\documentclass[epjCONF]{svjour}
\usepackage{graphics,graphicx}
\usepackage[varg]{txfonts} 
\usepackage[latin1]{inputenc}
\usepackage{amsfonts}
\session-title{Fusion11}
\begin{document}
\title{Microscopic Calculation of Heavy-Ion Potentials Based on TDHF}
\author{A.S. Umar\inst{1}\fnmsep\thanks{\email{umar@compsci.cas.vanderbilt.edu}} \and V.E. Oberacker\inst{1} \and J.A. Maruhn\inst{2} \and P.-G. Reinhard\inst{3}}
\institute{Department of Physics and Astronomy, Vanderbilt University, Nashville, Tennessee 37235, USA \and Institut f\"ur Theoretische Physik, Goethe-Universit\"at, D-60438 Frankfurt am Main, Germany \and Institut f\"ur Theoretische Physik, Universit\"at Erlangen, D-91054 Erlangen, Germany}
\abstract{
We discuss the implementation and results of a recently developed microscopic method for calculating
ion-ion interaction potentials and fusion cross-sections. The method uses the TDHF evolution to obtain
the instantaneous many-body collective state using a density constraint. The ion-ion potential  as well as the 
coordinate dependent mass are calculated from these states. The method fully accounts for the dynamical
processes present in the TDHF time-evolution and provides a parameter-free way of calculating fusion
cross-sections.
} 
\maketitle
\section{Introduction}
\label{intro}

The study of internuclear potentials for heavy-ion collisions is
of fundamental importance for calculating fusion cross-sections,
and for studying the formation of
superheavy elements and nuclei far from stability.
While asymptotically such potentials are determined from
Coulomb and centrifugal interactions, the short distance
behavior strongly depends on the nuclear surface properties
and the readjustments of the combined nuclear system,
resulting in potential pockets, which determine the
characteristics of the compound nuclear system.

Among the various approaches for calculating ion-ion
potentials are:
1) Phenomenological models such as the Bass model \cite{Ba80},
the proximity potential \cite{BR77}, and potentials obtained via
the double-folding method \cite{SL79,RO83a}. Some of these potentials
have been fitted to experimental fusion barrier heights
and have been remarkably successful in describing scattering data.
2) Semi-microscopic and full microscopic calculations such as the
macroscopic  -  microscopic method \cite{KN79,II05}, the 
asymmetric two-center shell-model \cite{MG72}, 
constrained Hartree-Fock (CHF) with a constraint on the quadrupole moment
or some other definition of the internuclear distance \cite{ZM76,BB04},
and other mean-field based calculations \cite{DN02,DP03,GB05}.

One common physical assumption used in many of the semi-microscopic
calculations is the use of the {\it frozen density} or the {\it sudden}
approximation. As the name suggests, in this approximation the nuclear densities
are unchanged during the computation of the ion-ion potential as a function
of the internuclear distance. On the other hand, the microscopic calculations
follow a minimum energy path and allow for the rearrangement of the nuclear
densities as the relevant collective parameter changes.
In this paper, we
shall call this the {\it static adiabatic approximation} since a real
adiabatic calculation would involve a fully dynamical calculation, thus
also including the effects of dynamical rearrangements.

Recently, we have developed a new method to extract ion-ion interaction potentials directly from
the TDHF time-evolution of the nuclear system.
In the DC-TDHF approach \cite{UO06b}
the TDHF time-evolution takes place with no restrictions.
At certain times during the evolution the instantaneous density is used to
perform a static Hartree-Fock minimization while holding the neutron and proton densities constrained
to be the corresponding instantaneous TDHF densities. In essence, this provides us with the
TDHF dynamical path in relation to the multi-dimensional static energy surface
of the combined nuclear system. In this approach
there is no need to introduce constraining operators which assume that the collective
motion is confined to the constrained phase space. In short, we have a self-organizing system which selects
its evolutionary path by itself following the microscopic dynamics.
Some of the effects naturally included in the DC-TDHF calculations are: neck formation, mass exchange,
internal excitations, deformation effects to all order, as well as the effect of nuclear alignment
for deformed systems.

\section{DC-TDHF Method}
\label{sec:2}
\subsection{Formalism}
\label{sec:2.1}

The concept of using density as a constraint for calculating collective states
from TDHF time-evolution was first introduced in Ref.~\cite{CR85}, and used
in calculating collective energy surfaces in connection with nuclear molecular
resonances in Ref.~\cite{US85}.

In this approach we assume that a collective state is characterized only by
density  $\rho$, and current $\mathbf{j}$. This state can be constructed
by solving the static Hartree-Fock equations
\begin{equation}
<\Phi_{\rho,\mathbf{j}}|a_h^{\dagger}a_p\hat{H}|\Phi_{\rho,\mathbf{j}}>=0\;,
\end{equation}
subject to constraints on
density and current
\begin{eqnarray*}
<\Phi_{\rho,\mathbf{j}}|\hat{\rho}(\mathbf{r})|\Phi_{\rho,\mathbf{j}}>&=&\rho(\mathbf{r},t) \\
<\Phi_{\rho,\mathbf{j}}|\hat{\jmath}(\mathbf{r})|\Phi_{\rho,\mathbf{j}}>&=&\mathbf{j}(\mathbf{r},t)\;.
\end{eqnarray*}
Choosing $\rho(\mathbf{r},t) $ and $\mathbf{j}(\mathbf{r},t)$ to be the instantaneous TDHF
density and current results in the lowest energy collective state corresponding to the
instantaneous TDHF state $|\Phi(t)>$, with the corresponding energy
\begin{equation}
E_{coll}(\rho(t),\mathbf{j}(t))=<\Phi_{\rho,\mathbf{j}}|\hat{H}|\Phi_{\rho,\mathbf{j}}>\;.
\end{equation}
This collective energy differs from the conserved TDHF energy only by the amount of
internal excitation present in the TDHF state, namely
\begin{equation}
E^{*}(t)=E_{TDHF} - E_{coll}(t)\;.
\end{equation}
However, in practical calculations the constraint on the current is difficult to implement
but we can define instead a static adiabatic collective state $|\Phi_{\rho}>$ subject to the
constraints
\begin{eqnarray*}
<\Phi_{\rho}|\hat{\rho}(\mathbf{r})|\Phi_{\rho}>&=&\rho(\mathbf{r},t) \\
<\Phi_{\rho}|\hat{\jmath}(\mathbf{r})|\Phi_{\rho}>&=&0\;.
\end{eqnarray*}
In terms of this state one can write the collective energy as
\begin{equation}
\label{eq:4}
E_{coll}=E_{kin}(\rho(t),\mathbf{j}(t))+E_{DC}(\rho(\mathbf{r},t))\;,
\end{equation}
where the density-constrained energy $E_{DC}$, and the collective kinetic
energy $E_{kin}$ are defined as
\begin{eqnarray*}
E_{DC}&=&<\Phi_{\rho}|\hat{H}|\Phi_{\rho}> \\
E_{kin}&\approx&\frac{\hbar^2}{2m}\int d^{3}r\; \mathbf{j}(t)^2/\rho(t)\;.
\end{eqnarray*}
From Eq.~\ref{eq:4} is is clear that the density-constrained energy
$E_{DC}$ plays the role of a collective potential. In fact this is
exactly the case except for the fact that it contains the binding
energies of the two colliding nuclei. One can thus define the ion-ion
potential as
\begin{equation}
V=E_{\mathrm{DC}}(\rho(\mathbf{r},t))-E_{A_{1}}-E_{A_{2}}\;,
\end{equation}
where  $E_{A_{1}}$ and $E_{A_{2}}$ are the binding energies of two nuclei
obtained from a static Hartree-Fock calculation with the same effective
interaction. For describing a collision of two nuclei one can label the
above potential with ion-ion separation distance $R(t)$ obtained during the
TDHF time-evolution. This ion-ion potential $V(R)$ is asymptotically correct
since at large initial separations it exactly reproduces $V_{Coulomb}(R_{max})$.
In addition to the ion-ion potential it is also possible to obtain coordinate
dependent mass parameters. One can compute the ``effective mass'' $M(R)$
using the conservation of energy
\begin{equation}
M(R)=\frac{2[E_{\mathrm{c.m.}}-V(R)]}{\dot{R}^{2}}\;,
\label{eq:mr}
\end{equation}
where the collective velocity $\dot{R}$ is directly obtained from the TDHF evolution and the potential
$V(R)$ from the density constraint calculations.

\subsection{Calculation of $R$}
In practice, TDHF runs are initialized with energies above the Coulomb barrier at some
large but finite separation. The two ions are boosted with velocities obtained by assuming
that the two nuclei arrive at this initial separation on a Coulomb trajectory.
Initially the nuclei are placed such that the center of mass is located at $x=y=z=0$,
and the $x-z$ plane represents the collision plane.
During the TDHF dynamics the ion-ion separation distance is obtained by constructing a dividing plane between the
two centers and calculating the center of the densities on the left and right halves of this
dividing plane. The coordinate $R$ is the difference between the two centers.
The dividing plane is determined by finding the point at which the tails of the two
densities intersect each other along the $x$-axis. However, this procedure
starts to fail after a substantial overlap is reached.
Instead, one can define the 
ion-ion separation as $R=R_0\sqrt{|Q_{20}|}$, where $Q_{20}$ is the mass quadrupole moment
for the entire system, calculated by diagonalizing the quadrupole tensor to obtain the quadrupole moment
along the principal axis,
and $R_0$ is a scale factor determined to give the correct initial
separation distance at the start of the calculations. Calculating $R$ this way yields
numerically identical results to the previous procedure until that procedure begins to fail
and continues smoothly after that point.

\subsection{Fusion cross-section}
We now outline the calculation of the total fusion cross section using a
coordinate-dependent mass $M(R)$ and potential $V(R)$. Starting from the quantized Hamiltonian
\begin{equation}
H(R,\hat{P})=\frac{1}{2} \left[ M(R)^{-{\frac{1}{2}}} \hat{P} M(R)^{-{\frac{1}{2}}} \hat{P} \right] + V(R) \;.
\label{eq:Ham2}
\end{equation}
with the momentum operator $\hat{P} = -i \hbar d/dR$,
the total fusion cross cross-section
\begin{equation}
\sigma_f = \frac{\pi}{k^2} \sum_{L=0}^{\infty} (2L+1) T_L\;,
\label{eq:sigfus}
\end{equation}
can be obtained by calculating the potential barrier penetrabilities $T_L$
from the Schr\"odinger equation for the relative motion coordinate $R$
using the Hamiltonian~(\ref{eq:Ham2}) with an additional centrifugal potential
\begin{equation}
\left [ H(R,\hat{P}) + \frac{\hbar^2 L(L+1)}{2 M(R) R^2}
 - E_\mathrm{c.m.} \right] \psi_L(R) = 0 \;.
\label{eq:Schroed1}
\end{equation}

Alternatively, instead of solving the Schr\"odinger equation with coordinate dependent
mass parameter $M(R)$ for the heavy-ion potential $V(R)$, we can instead use the constant
reduced mass $\mu$ and transfer the coordinate-dependence of the mass to a scaled
potential $U(\bar{R})$ using the well known coordinate scale transformation.
\begin{equation}
d\bar{R}=\left(\frac{M(R)}{\mu}\right)^{\frac{1}{2}}dR\;.
\label{eq:mrbar}
\end{equation}
Integration of Eq.~(\ref{eq:mrbar}) yields
\begin{equation}
\bar{R}= f(R) \ \ \ \Longleftrightarrow \ \ \ R=f^{-1}(\bar{R})\;.
\label{eq:rbar}
\end{equation}
As a result of this point transformation, both the classical
Hamilton function and the corresponding quantum mechanical
Hamiltonian, Eq.~(\ref{eq:Ham2}), now assume the form
\begin{equation}
H(\bar{R},\bar{P})=\frac{\bar{P}^2}{2 \mu} + U(\bar{R}) \;,
\label{eq:Ham3}
\end{equation}
and the scaled heavy-ion potential is given by the expression
\begin{equation}
U(\bar{R}) = V(R) = V(f^{-1}(\bar{R})) \;.
\label{eq:Urbar}
\end{equation}

The fusion barrier penetrabilities $T_L(E_{\mathrm{c.m.}})$
are obtained by numerical integration of the two-body Schr\"odinger equation
using the {\it incoming wave boundary condition} (IWBC) method.
IWBC assumes that once the minimum of the potential is reached fusion will
occur. In practice, the Schr\"odinger equation is integrated from the potential
minimum, $R_\mathrm{min}$, where only an incoming wave is assumed, to a large asymptotic distance,
where it is matched to incoming and outgoing Coulomb wavefunctions. The barrier
penetration factor, $T_L(E_{\mathrm{c.m.}})$ is the ratio of the
incoming flux at $R_\mathrm{min}$ to the incoming Coulomb flux at large distance.
Here, we implement the IWBC method exactly as it is
formulated for the coupled-channel code CCFULL described in Ref.~\cite{HR99}.
This gives us a consistent way for calculating cross-sections at above and below
the barrier energies. 

\subsection{ Fusion with alignment}
\label{sec:align}

In the case of one or both of the reaction partners being deformed one has to
incorporate the nuclear alignment into the evolution of the heavy-ion collision dynamics.
This is done in two steps~\cite{UO06c}:
a) A dynamical Coulomb alignment calculation to determine the probability that a given nuclear orientation
occurs at the distance $R(t_0)$, where the TDHF run is initialized.
The alignment generally results from multiple {\it E2}/{\it E4} Coulomb excitation
of the ground state rotational band.
The distance $R(t_0)$ is chosen such
that the nuclei only interact via the Coulomb interaction.
b) A TDHF calculation, starting at this finite internuclear distance $R(t_0)$, for
a fixed initial orientation of the deformed nucleus.
Since the experiments are usually done with unpolarized beams, in
a full quantum mechanical calculation one would have to average over discrete
quantum mechanical rotational bands. In the classical limit, this
corresponds to averaging over orientation angles.
In the case of one spherical nucleus and one deformed reaction partner, the total fusion cross
section is given by an integral over all orientation (Euler) angles, with solid
angle element $d\Omega=2\pi sin\beta d\beta$
\begin{equation}
\label{eq:fusion}
\sigma(E_{\mathrm{c.m.}}) = \int d\Omega\; \frac{dP}{d\Omega}\; \sigma(E_{\mathrm{c.m.}},\Omega)\;,
\end{equation}
where $dP/d\Omega$ represents the alignment probability and $\sigma(E_{\mathrm{c.m.}},\Omega)$
is the fusion cross section associated with a particular alignment calculated using the ion-ion
potential, $V(R,\beta)$, obtained from the TDHF collision for which the deformed partner
is initialized with angle $\beta$ with respect to the collision axis. Details for the most general
case is given in Refs.~\cite{UO06c,UO08a}.

\section{Application}
\label{sec:3}
In this section we will give selected examples of the DC-TDHF
method for calculating fusion cross-sections.
Calculations were done in a  3-D Cartesian box large enough to avoid any initialization or
box boundary effects.
We have used the full Skyrme force (SLy4)~\cite{CB98}.
The numerical accuracy of the static binding energies and the deviation
from the point Coulomb energy in the initial state of the collision dynamics is on the order
of $50-150$~keV. We have performed density constraint calculations at every $10-20$~fm/c interval.
The accuracy of the density constraint calculations are
commensurate with the accuracy of the static calculations.
\begin{figure}[!htb]
\begin{center}
\resizebox{0.90\columnwidth}{!}{ \includegraphics*{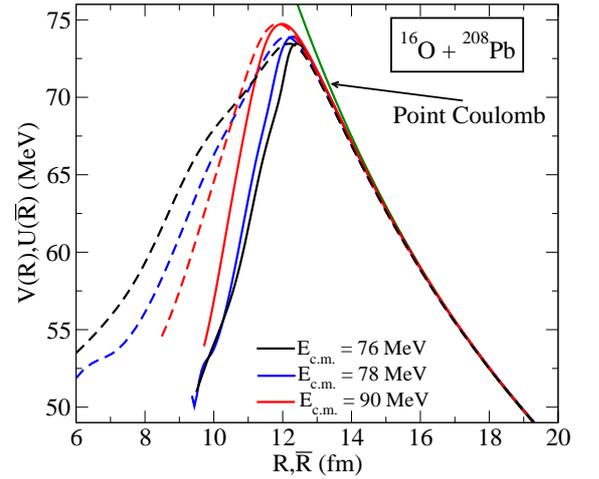} }
\caption{\label{fig1} (Colour on-line) Potential barriers obtained from density constraint
TDHF calculations at three different energies. The three dashed curves
correspond to the transformed potential using coordinate dependent masses.}
\end{center}
\end{figure}
\begin{figure}[!htb]
\begin{center}
\resizebox{0.90\columnwidth}{!}{ \includegraphics*{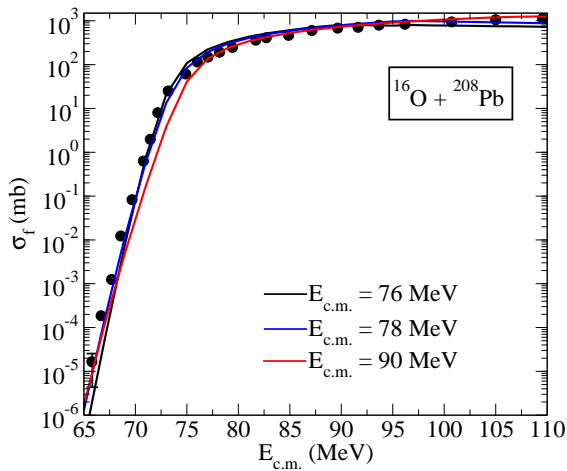} }
\caption{\label{fig2} (Colour on-line)  Total fusion cross section as a function of c.m. energy using the potentials of
Fig.~\protect\ref{fig1}}
\end{center}
\end{figure}

\subsection{Spherical system}
\label{sec:3.1}

The DC-TDHF method is expected to do best for nuclei that are well described by the
Skyrme HF calculations. One such reaction is the fusion of $^{16}$O+$^{208}$Pb system.
In Fig.~\ref{fig1} we show an example of microscopic potentials for the $^{16}$O+$^{208}$Pb
system at three difference center-of-mass energies~\cite{UO09a}. The dashed curves are the
corresponding potentials transformed via the microscopically calculated effective mass, $M(R)$.
We observe that all of the scaled barriers give a very good description of the
fusion cross-section at higher energies suggesting these cross-sections are
primarily determined by the barrier properties in the vicinity of the barrier
peak, whereas for the extreme sub-barrier cross-sections are influenced by what
happens in the inner part of the barrier and here the dynamics and
consequently the coordinate dependent mass
becomes very important (see Fig.~\ref{fig2}). 
Specifically, we can see from  Fig.~\ref{fig1} that as the c.m. energy is increased
the ion-ion potential peak increases but the inner part of the barrier becomes
narrower. This is due to the fact that for high energies the system does not
have enough time for rearrangements in the density to occur and the barrier
approaches the frozen-density limit. However, at lower energies substantial
density rearrangements occur which modifies the inner part of the barrier.
This modification is important for fusion cross-sections are deep sub-barrier
energies.

\subsection{Deformed systems}
\label{sec:3.2}

\begin{figure}[!htb]
\begin{center}
\resizebox{0.9\columnwidth}{!}{ \includegraphics*{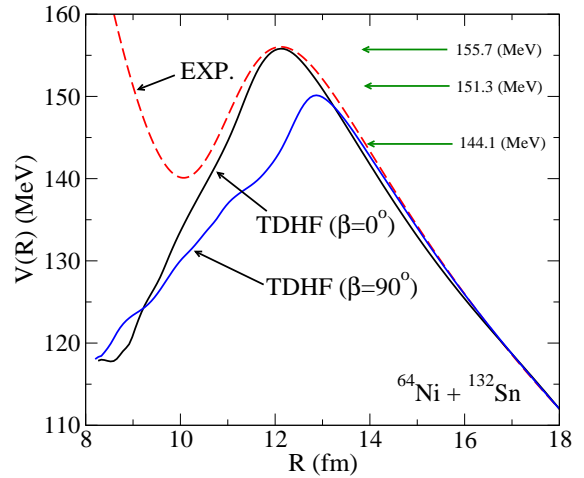} }
\caption{\label{fig3} (Colour on-line)  Potential barriers, $V(R,\beta)$, obtained from
DC-TDHF calculations for the $^{64}$Ni+$^{132}$Sn system. Angle $\beta$ indicates different
orientations of the deformed $^{64}$Ni nucleus in $\Delta\beta=10^{\circ}$ intervals. Also shown are
the experimental energies.}
\end{center}
\end{figure}

\begin{figure}[!htb]
\begin{center}
\resizebox{0.9\columnwidth}{!}{ \includegraphics*{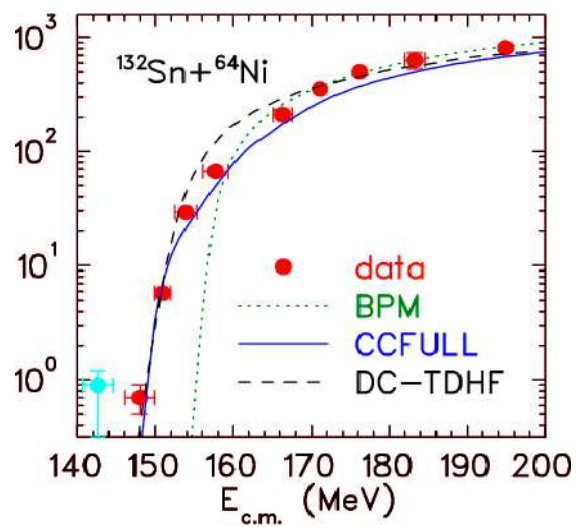} }
\caption{\label{fig4} (Colour on-line) Total fusion cross section as a function of $E_{\mathrm{c.m.}}$.
Shown are the experimental data (filled circles), the latest coupled-channel calculations~\cite{Li07}
including neutron transfer (blue solid curve), and the DC-TDHF cross sections  (dashed curve).}
\end{center}
\end{figure}

The collision of the $^{64}$Ni+$^{132}$Sn system represents a good example
of a collision involving a deformed (oblate) nucleus, $^{64}$Ni and a neutron rich nucleus. 
Fusion cross-sections for this system have been experimentally measured \cite{Li07}  and initially
a significant discrepancy was observed with standard coupled-channel calculations.
We have used the DC-TDHF method to study this system \cite{UO06d,UO07a}.
The ion-ion potentials corresponding to two extreme orientations of the $^{64}$Ni nucleus
are shown in Fig.~\ref{fig3} as well as an empirical barrier used in barrier penetration calculations
in Ref.~\cite{Li07}. Two important points are observed from this plot. The first is the strong
dependence of the barrier height and location on the alignment of the deformed nucleus.
We also see that the empirical barrier is very close to the equatorial orientation, which is 
closer to the assumption of spherical nuclei. The accuracy of our result with no parameters
or normalization is impressive. The second point has to do with the meaning of \textit{sub-barrier};
as seen from Fig.~\ref{fig3}, while the experimental energies appear to be all sub-barrier with
respect to the spherical barrier, two of them are above the barrier with respect to the $\beta=90^{o}$
potential barrier and the third faces a considerably narrower barrier. This explains the anomalous
observation of enhanced fusion at these energies.
Fig.~\ref{fig4} shows the experimental and theoretical fusion cross-sections calculated with different
methods. The coupled-channel calculations are modified to include multiple neutron transfer. As we
see again the DC-TDHF results reproduce the fusion cross-sections reasonably well.

\subsection{Superheavy systems}
\label{sec:3.3}

Ion-ion interaction potentials calculated using DC-TDHF correspond to the
configuration attained during a particular TDHF collision. For light and
medium mass systems as well as heavier systems for which fusion is
the dominant reaction product, DC-TDHF
gives the fusion barrier with an appreciable but relatively small energy dependence.
On the other hand, for reactions leading to superheavy systems fusion is
not the dominant channel at barrier top energies. Instead the system sticks
in some dinuclear configuration with possible break-up after exchanging a
few nucleons.
For this reason the energy dependence of the DC-TDHF interaction barriers
for these systems is not just due to the dynamical effects for the same final
configuration but actually represent different final configurations.
For the same reasons calculations presented here can only address the
capture cross-section for these systems since the long-time evolution to 
complete fusion or break-up is beyond the scope of
TDHF due to the absence of quantum decay processes and transitions.
\begin{figure}[!htb]
\resizebox{0.9\columnwidth}{!}{ \includegraphics*{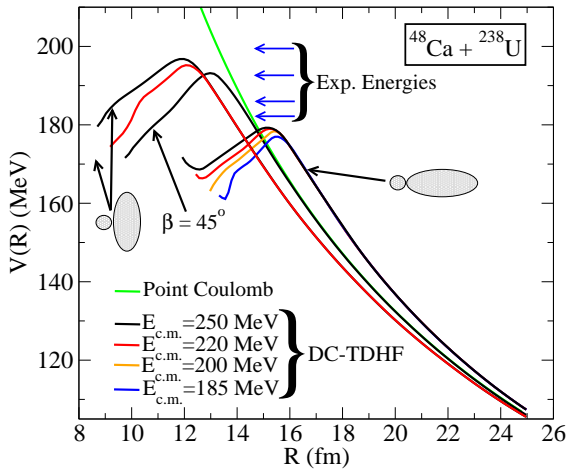} }
\caption{\label{fig5} (colour on-line)
Potential barriers, $V(R)$,  for the $^{48}$Ca+$^{238}$U system obtained from
DC-TDHF calculations as a function of $E_\mathrm{c.m.}$ energy and
for selected orientation angles $\beta$ of the $^{238}$U nucleus. Also, shown are the experimental c.m. energies.}
\end{figure}

As an example of superheavy formation from a hot-fusion reaction we have studied the $^{48}$Ca+$^{238}$U system \cite{UO10a}.
Hartree-Fock (HF) calculations produce a  spherical $^{48}$Ca nucleus, whereas $^{238}$U has a large axial deformation.
The large deformation of $^{238}$U is expected to strongly influence the interaction barriers for this system.
This is shown in Fig.~\ref{fig5}, which shows the interaction barriers, $V(R)$,
calculated using the DC-TDHF method as a function of c.m. energy and for three different
orientations of the $^{238}$U nucleus. The alignment angle $\beta$ is the angle
between the symmetry axis of the $^{238}$U nucleus and the collision axis.
Also shown in Fig.~\ref{fig5} is the point Coulomb potential corresponding to this collision.
The deviations from the point Coulomb potential at large $R$ values are due to the deformation
of the $^{238}$U nucleus.
We first notice that the barriers corresponding to the polar orientation ($\beta=0^{o}$) of the
$^{238}$U nucleus are much lower and peak at larger ion-ion separation distance $R$.
On the other hand, the barriers corresponding to the equatorial orientation of $^{238}$U ($\beta=90^{o}$) are
much higher and peak at smaller $R$ values. For the intermediate values of $\beta$ the barriers
rise rapidly as we increase the orientation angle from $\beta=0^{o}$, as can be seen for $\beta=45^{o}$.
The rise in the barrier height as a function of increasing $\beta$ values is not linear but seems to
rise more rapidly for smaller $\beta$ values.
We also see that for lower energies central collisions with polar orientation of $^{238}$U are the only orientations which result in
the sticking of the two nuclei, while the equatorial orientations of $^{238}$U result in a deep-inelastic
collision.
Also, shown in Fig.~\ref{fig5} are the experimental energies~\cite{Og07} for this reaction.
We observe that all of the experimental energies are above the barriers obtained for the
polar alignment of the $^{238}$U nucleus.
\begin{figure}[!htb]
\resizebox{0.9\columnwidth}{!}{ \includegraphics*{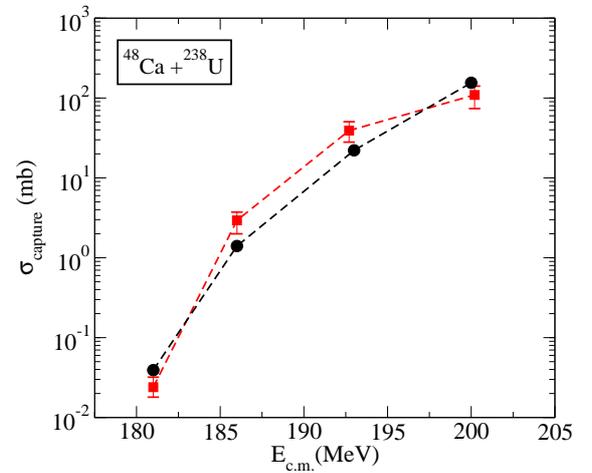} }
\caption{\label{fig6} (colour on-line) Capture cross-sections for the $^{48}$Ca+$^{238}$U system
as a function of $E_\mathrm{c.m.}$ energy (black circles). Also, shown are the experimental cross-sections
(red squares)~\cite{Og07}.}
\end{figure}
For the calculation of the capture cross-section we need to average over all possible
alignments of the $^{238}$U nucleus as indicated by Eq.~(\ref{eq:fusion}).
Due to the relatively small charge of the $^{48}$Ca nucleus the alignment probability 
$dP/d\Omega$ of Eq.~(\ref{eq:fusion}) is in the range $0.46-0.52$
and does not vary appreciably with energy.
In Fig.~\ref{fig6} we show the capture cross-sections for the $^{48}$Ca+$^{238}$U system
as a function of $E_\mathrm{c.m.}$ energy (black circles). Also, shown are the experimental cross-sections
(red squares)~\cite{Og07}.
One important fact to notice in the
cross-section formula given in Eq.~(\ref{eq:fusion}) is that 
the cross-section is multiplied by the $\sin(\beta)$ factor, which renders the contribution
originating from the lowest barriers at small values of $\beta$ to be very small.
\begin{figure}[!htb]
\resizebox{0.9\columnwidth}{!}{ \includegraphics*{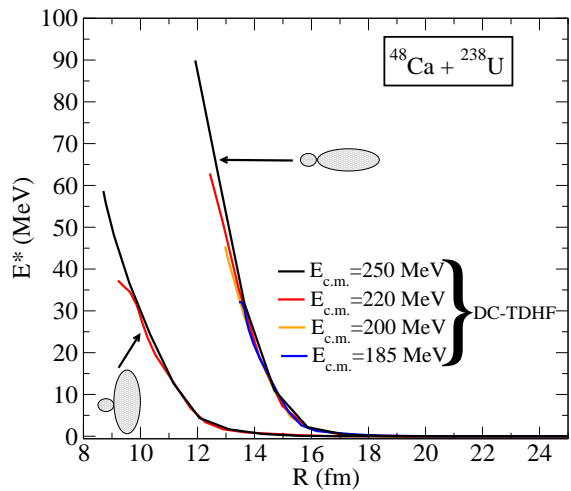} }
\caption{\label{fig7} (colour on-line)
Excitation energy, $E^{*}(R)$, in a central collision for various
$E_\mathrm{c.m.}$ energies and for alignment angles angles $\beta=0^{o}$ and $\beta=90^{o}$ of the $^{238}$U nucleus.}
\end{figure}

In Fig.~\ref{fig7} we also show the excitation energy $E^{*}(R)$ as a function of c.m. energy
and for two alignment angles ($\beta=0^{o}$ and $\beta=90^{o}$ ) of the $^{238}$U nucleus.
The excitation energy curves start at zero excitation when the two nuclei are far apart,
which also provides a test for the numerical accuracy of the calculation.
We note that the system is
excited much earlier during the collision process for the polar alignment of the $^{238}$U nucleus
and has a higher excitation than the corresponding collision for the equatorial orientation.
Only two curves are shown for the equatorial collision since at lower energies we have
deep-inelastic collisions for this alignment.
We note that the highest point reached for these excitation curves is chosen
to be the point where the nuclei almost come to a stop inside the barrier, which corresponds to a nearly zero collective
kinetic energy.
Since this is determined during the initial phase of the collision the dinuclear system is not in thermal equilibrium.
However, the system essentially oscillates about this point.
For energies for which the collision outcome is capture
this would be the excitation energy at the capture point.

\section{Conclusions}
\label{sec:4}

In summary, we have used the fully microscopic DC-TDHF method to obtain ion-ion
interaction barriers for calculating fusion and capture cross-sections.
The standard, parameter free approach of DC-TDHF yields potential barriers that can
accurately reproduce the fusion cross-sections. The DC-TDHF approach has now been applied
to study a number of systems with very promising results. 
This further elucidates the point that for the proper description of fusion cross-sections
dynamical effects such as neck formation and mass transfer
must be included to modify the inner part of the potential barrier. Also interesting
are the implications of the energy dependent barriers.
These findings underscore the fact that
additional modifications needed for phenomenological methods to explain the fusion
cross-sections  may largely be due to the inadequacy
of the approximations made in treating the nuclear dynamics.

This work has been supported by the U.S. Department of Energy under grant No.
DE-FG02-96ER40963 with Vanderbilt University, and by the German BMBF
under contract Nos. 06FY9086 and 06ER142D.

\end{document}